\documentclass
[
showpacs,
aps,
preprint,
amsmath,
groupedaddress
]
{revtex4-1}
\usepackage{physics,graphicx,color}
\usepackage{cleveref}
\crefname{equation}{Eqn.}{Eqns.}
\begin{document}


\title{Non-Equilibrium Modeling of the Fe XVII 3C/3D ratio for an Intense X-ray Free Electron Laser}


\author{Y.~Li}
 \email[]{yzl0060@auburn.edu}
\author{M.~Fogle}
\author{S.~D.~Loch}
 \email[]{loch@physics.auburn.edu}
\affiliation{Department of Physics, Auburn University, Auburn AL 36849, USA}

\author{C.~P.~Ballance}
\affiliation{Queen's University, Belfast, Belfast, BT7 1NN, UK}

\author{C.~J.~Fontes}
\affiliation{Los Alamos National Laboratory, Los Alamos, NM 87545, USA}


\date{\today}

\begin{abstract}
We present a review of two methods used to model recent LCLS experimental results for the 3C/3D line intensity ratio of
Fe XVII \cite{Bernitt2012}, 
the time-dependent collisional-radiative method and the density-matrix approach. These are described and
applied to a two-level atomic system excited by an X-ray free electron laser. 
A range of pulse parameters is explored and the effects on the predicted Fe XVII 3C and 3D line intensity ratio are calculated. 
In order to investigate the behavior of the predicted line intensity ratio, a particular
pair of A-values for the 3C and 3D transitions was chosen (2.22~$\times$~10$^{13}$~s$^{-1}$ and 6.02~$\times$~10$^{12}$~s$^{-1}$ for the 3C and 3D, respectively), but our conclusions are independent of the precise values.
We also reaffirm the conclusions from Oreshkina et al.~\cite{Oreshkina2014,Oreshkina2015}: the non-linear effects in the density 
matrix are important and the reduction in the Fe XVII 3C/3D line intensity ratio is sensitive to the laser pulse parameters, namely pulse duration, pulse intensity, and laser bandwidth. 
It is also shown that for both models the lowering of the 3C/3D line intensity ratio below the expected time-independent oscillator strength ratio has a significant contribution due to the emission from the plasma after the laser pulse has left the plasma volume. 
Laser intensities above $\sim 1\times 10^{12}$ W/cm$^{2}$ are required for a reduction in the 3C/3D line intensity ratio below the expected time independent oscillator strength ratio. 

\end{abstract}

\pacs{32.70.Cs}

\maketitle





\section{Introduction}
Spectral emission from Fe XVII can be used as a valuable plasma diagnostic for both laboratory and astrophysical plasmas~\cite{Phillips1997,Brinkman2000}.  The ratio of the 3C line intensity (transition 2p$^5$ 3d ($^1$P$_1$)~$\rightarrow$~2p$^6$ ($^1$S$_0$)) to the 3D line intensity (transition 2p$^5$ 3d ($^3$D$_1$)~$\rightarrow$~2p$^6$ ($^1$S$_0$)) is sensitive to the plasma electron temperature and has been the focus of much attention in the literature.   During the history of disagreement between theory and observation for this line ratio, a number of underlying effects were found to be important, including blending with an inner shell satellite line of Fe XVI~\cite{Brown2001} and radiative cascades~\cite{Loch2006,Chen2007}. In addition, Gu~\cite{Gu2009} explored the possibility that insufficient configuration-interaction was included in the atomic structure calculations leading to unconverged oscillator strengths.  He then used an approximate method to account for this lack of convergence to modify the atomic collision data used in Fe XVII spectral modeling.  A full discussion of the comparison of theory and experiment for this line ratio is outside of the scope of this article.  Brown~\cite{Brown2008} presents a review of measurement results and Brown and Beiersdorfer~\cite{Brown2012} show a useful summary of the discrepancies and the effects that have been investigated. The focus of this article is on the analysis of a recent experiment using an X-ray Free Electron Laser (XFEL) that sought to identify the source of the aforementioned discrepancies~\cite{Bernitt2012}.  

Bernitt et al.~\cite{Bernitt2012} used an intense XFEL at the Linac Coherent Light Source (LCLS), employing the laser to excite Fe$^{16+}$ ions in an Electron Beam Ion Trap (EBIT).  The laser has a narrow bandwidth and was tuned to only populate the upper level of either the 3C or the 3D transition.  In this two-level setup, the observed 3C/3D line intensity ratio was expected to be the same as the 3C/3D oscillator strength ratio, and any differences could be interpreted as an indicator of deficiencies in the current atomic structure calculations for Fe$^{16+}$.  The experiment resulted in a much lower 3C/3D line intensity ratio (2.61 $\pm$ 0.13) than the previously calculated oscillator strength ratios ($\sim 3.5$ or higher).  It was also pointed out that the 3C/3D oscillator strength ratio is only slowly converging with the increasing size of the configuration-interaction expansion included in the theoretical calculations.  The most complete theoretical calculations all produced oscillator strength ratios significantly larger (3.5~\cite{Gu2009}, 3.54~\cite{Chen2007} 3.42~\cite{Safronova2001}, and 3.49~\cite{Bernitt2012}) than the observed line intensity ratio from the LCLS experiment.

To investigate the unexpectedly low 3C/3D line intensity ratio observed from the XFEL experiment, two approaches were adopted.  A density matrix (D-M) approach, first employed by Oreshkina et al.~\cite{Oreshkina2014,Oreshkina2015} and reproduced in this paper, showed that the 3C/3D line intensity ratio can be reduced below the expected oscillator strength ratio for sufficiently intense laser pulses and that the reduction is sensitive to certain laser pulse parameters (intensity, duration and bandwidth).
Alternatively, Loch et al.~\cite{Loch2015} used a collisional-radiative (C-R) method and showed that the spectral emission from the plasma after the laser pulse has left the plasma volume makes a strong contribution to the lowering of the 3C/3D line intensity ratio. 

In this paper both the C-R and the D-M approaches are summarized. The D-M method is preferred for intense laser fields, due to the possible non-linear response of the excited populations with laser intensity and the phase of the electric field.  In Section~\ref{sec:theory} both theoretical methods are described, in Section~\ref{sec:results} the results using each method are shown, and in Section~\ref{sec:conclusion} some discussion and possible future directions are presented.

\section{Theory}
\label{sec:theory}

\subsection{C-R Method}
The C-R method is used widely in laboratory and astrophysical plasma modeling. 
This approach takes into account all of the 
atomic process in a rate matrix, from which the steady-state and time-dependent populations can be evaluated.  
The laser bandwidth in the LCLS experiment was sufficiently narrow to ensure that only one transition in Fe$^{16+}$ could be excited at a time, 
thus this could be treated as a two-level system. 
For both the 3C and 3D lines, the only populating mechanism for the excited 
state is photo-absorption from the ground level and the only associated depopulating mechanisms 
are stimulated emission (sometimes referred to as the interacting process) and spontaneous emission (the non-interacting
process). The time-dependent population density for the 
excited state $N_{e}$ and ground state $N_{g}$ can be evaluated (see, e.g., 
Bethe and Jackiw~\cite{Bethe1968} page 204--205):

\begin{equation}\label{equ-cr-e}
    \frac{dN_{e}}{dt} = N_{g}(t) \rho ( \omega _{0} , t )B_{g \rightarrow e} - N_{e} (t) (A_{e\rightarrow g} + \rho ( \omega _{0} , t ) B_{e \rightarrow g})
\end{equation}

\begin{equation}\label{equ-cr-g}
    \frac{dN_{g}}{dt} = - N_{g}(t) \rho ( \omega _{0} , t )B_{g \rightarrow e} + N_{e} (t) (A_{e\rightarrow g} + \rho ( \omega _{0} , t ) B_{e \rightarrow g})
\end{equation}

\noindent where $B_{g \rightarrow e}$, $B_{e \rightarrow g}$,  $A_{e \rightarrow g}$ are the 
Einstein photo-absorption, stimulated emission, and spontaneous emission coefficients, respectively.  $\omega_{0}$ is the angular frequency for the transition between the two levels.  
These can be evaluated from
atomic structure calculations. $\rho$ is the radiation field density (J/m$^{3}$/Hz) and can be determined from 
the laser parameters.  In the D-M approach the laser intensity $I$~(W/cm$^{2}$) is used, so it is beneficial to be able to convert between the two representations via
$ \rho=I/(c \cdot \delta \nu)$.  Here $c$ is the speed of light and $\delta \nu$ is the bandwidth of the laser (e.g. $I$ = 10$^{10}$ W/cm$^{2}$~$\rightarrow$~$\rho = 1.10\times10^{-9}$ J/m$^{3}$/Hz).
In order to solve the time-dependent~\cref{equ-cr-e,equ-cr-g}, the matrix form is used:

\begin{equation}\label{equ-cr-matrix}
\begin{bmatrix}
dN_{g}/dt \\
dN_{e}/dt
\end{bmatrix}
=
\begin{bmatrix}
    -\rho ( \omega _{0} , t )B_{g \rightarrow e}&  A_{e\rightarrow g} + \rho ( \omega _{0} , t ) B_{e \rightarrow g} \\
     \rho ( \omega _{0} , t )B_{g \rightarrow e}& -(A_{e\rightarrow g} + \rho ( \omega _{0} , t ) B_{e \rightarrow g})
\end{bmatrix}
\begin{bmatrix}
N_{g} \\
N_{e}
\end{bmatrix}
.
\end{equation}

Initially, one hundred percent of the population is fixed to be in the ground state. Thus, 
the initial normalized population vector is $\begin{bmatrix} 1& 0 \end{bmatrix}$$^{T}$, where the superscript indicates the transpose. The excited state population $N_e(t)$ is evaulated for a given $\rho(t)$ using~\cref{equ-cr-matrix}. This can then be used to determine the photon emission for the time during which the laser pulse is in the plasma volume.  Note that while stimulated emission is included in the modeling of the excited population density (see~\cref{equ-cr-matrix}), these photons are not counted in the predicted line intensity (see~\cref{equ-cr-pho}) since the stimulated emission photons 
are emitted in the direction of the laser beam and not towards the detector. 
After the laser pulse has left the plasma volume, there will be a number of electrons left in the excited
state. All of these will decay via spontaneous emission before the next laser pulse.  Thus, there is a second contribution to the line emission with each of these excited state electrons producing one photon.
That is, the total photon energy detected in the spectral line will be proportional to:

\begin{equation}\label{equ-cr-pho}
    I^{photon}_{e\rightarrow g} = \hbar\omega_{0}A_{e\rightarrow g}\int_{0}^{T} N_{e}(t)  dt + \hbar\omega_{0}N_{e} (T).
\end{equation}

The first term on the right hand side represents the emission during the time, indicated by $T$, that the laser pulse is interacting with the EBIT plasma and the second term represents the contribution to the emission from the plasma after the laser pulse has passed. 
Clearly the laser pulse temporal profile is a key factor in evaluating the time-dependent excited populations. 
Various envelopes for $\rho(t)$ have been considered and will be shown later in this article.


\subsection{Density-Matrix Method}
The D-M approach is a different formalism compared to the C-R approach.
For a two-level system in a stationary state, the ground and excited levels have eigenvalues $\hbar \omega_{g}$ and $\hbar \omega_{e}$, 
and wave functions $\Psi_{g}(\vec{r})$ and $\Psi_{e}(\vec{r})$ in the Heisenberg picture.
The total wave function of the system can be expressed as: 
\begin{equation}\label{equ-wf-s-t}
    \Psi(\vec{r},t) = C_{g} (t) \Psi_{g}(\vec{r}) + C_{e}(t) \Psi_{e}(\vec{r}).
\end{equation} 
The density operator is defined as $\rho$=$\ket{\Psi}\bra{\Psi}$, which has the form

\begin{equation}\label{equ-dm-ele}
    \rho
    =
\begin{bmatrix}
    \bra{g} \rho \ket{g} & \bra{g} \rho \ket{e} \\
    \bra{e} \rho \ket{g} & \bra{e} \rho \ket{e}
\end{bmatrix}
    =
\begin{bmatrix}
    \frac{N_{g}}{N_{g}+N_{e}} & C_{g}C_{e}^{*}\\
    C_{e}C_{g}^{*} & \frac{N_{e}}{N_{g}+N_{e}}
\end{bmatrix}
\end{equation}
\noindent where $\frac{N_{g}}{N_{g}+N_{e}}$ and $\frac{N_{e}}{N_{g}+N_{e}}$ are referred to as the populations and the products $C_{g}C_{e}^{*}$ and $C_{e}C_{g}^{*}$ are referred to as the coherence terms. 
For systems interacting with a laser, the Hamiltonian of the system can be written as: 
\begin{equation}\label{equ-ham-t}
    H=H_{S} + H_{I},
\end{equation} 
where the first term represents the stationary Hamiltonian given by
\begin{equation}\label{equ-ham-s}
    H_{S} = \hbar \omega_{g}\ket{\Psi_{g}}\bra{\Psi_{g}} + \hbar \omega_{e}\ket{\Psi_{e}}\bra{\Psi_{e}}
\end{equation}
and the second term represents the interaction Hamiltonian
\begin{equation}\label{equ-ham-i-1}
    H_{I} = -\vec{D}\cdot\vec{E} ,
\end{equation}
where $\vec{D}$ is the dipole moment and $\vec{E}$ is the radiation field. For a linearly polarized 
electric field along the $z$-axis, it can be written as $\vec{E}$ = $E_{0}(t)\cos(\omega_{L} t + \psi(t))\vec{z}$, 
where $E_{0}(t)$ is the electric field amplitude.  $E_{0}(t)$  can be determined from the radiation field intensity $I$ via $I=\frac{1}{2}c\epsilon_{0}nE_{0}$, where $c$ is the speed of the light, $\epsilon_{0}$ is the electric permittivity of free space, and $n$ is the refractive index of the medium.  $\omega_L$ is the angular frequency of the laser and $\psi(t)$ is the time-dependent phase of the laser field.
Using the rotating wave approximation (RWA), the interaction Hamiltonian can be further expanded as 
\begin{equation}\label{equ-ham-i-2}
    H_{I} =  -\frac{\hbar\Omega^{*}}{2}\ket{e}\bra{g} e^{-i\omega_{L} t}-\frac{\hbar\Omega}{2}\ket{g}\bra{e} e^{i\omega_{L} t}, 
\end{equation}
where $\Omega$ is the Rabi-frequency given by $\Omega = E_{0}(t)D_{eg}e^{i\psi (t)}/\hbar$, with $D_{eg} = e\bra{e}\hat z\ket{g}$ being the eletric dipole matrix element. 
The density operator $\rho$ is governed by the equation:
\begin{equation}\label{equ-drhodt}
    \frac{d\rho}{dt} = \frac{1}{i\hbar}[H,\rho] + \Lambda\rho
\end{equation}
where $\Lambda$ is the decay term due to spontaneous emission. 
From \cref{equ-drhodt}, one can show that:
\begin{subequations}\label{equ-dm-bloch1}
\begin{align}
     \frac{d\rho_{gg}}{dt}& = \Gamma \rho_{ee} - \frac{i\Omega^{*}}{2}e^{-i\omega_{L} t}\rho_{ge} + \frac{i\Omega}{2}e^{i\omega_{L} t}\rho_{eg}\label{equ-dm-bloch1-gg}\\
     \frac{d\rho_{ee}}{dt}& = -\Gamma \rho_{ee} + \frac{i\Omega^{*}}{2}e^{-i\omega_{L} t}\rho_{ge} - \frac{i\Omega}{2}e^{i\omega_{L} t}\rho_{eg}\label{equ-dm-bloch1-ee}\\
     \frac{d\rho_{eg}}{dt}& = \frac{i\Omega^{*}}{2}e^{-i\omega_{L} t}\rho_{gg} - \frac{i\Omega^{*}}{2}e^{-i\omega_{L} t}\rho_{ee} -(i\omega_{0} + \frac{\Gamma}{2}) \rho_{eg}\label{equ-dm-bloch1-eg}.
\end{align}
\end{subequations}
By using $\rho_{ge}=\rho_{ge}^{*}$, it is straightforward to get the expression for $\frac{d\rho_{ge}}{dt}$. 
By defining a new variable $\tilde \rho = e^{i\omega_{L} t} \rho$ and a detuning parameter $\Delta = \omega_{L} - \omega_{0}$, \cref{equ-dm-bloch1-gg,equ-dm-bloch1-ee,equ-dm-bloch1-eg} can be rewritten as follows:
\begin{subequations}\label{equ-dm-bloch2}
\begin{align}
     \frac{d\rho_{gg}}{dt}& = \Gamma \rho_{ee} - \frac{i\Omega^{*}}{2} \tilde \rho_{ge} + \frac{i\Omega}{2} \tilde \rho_{eg}\label{equ-dm-bloch2-gg}\\
     \frac{d\rho_{ee}}{dt}& = -\Gamma \rho_{ee} + \frac{i\Omega^{*}}{2} \tilde \rho_{ge} - \frac{i\Omega}{2} \tilde \rho_{eg}\label{equ-dm-bloch2-ee}\\
     \frac{d \tilde \rho_{eg}}{dt}& = \frac{i\Omega^{*}}{2} \rho_{gg} - \frac{i\Omega^{*}}{2} \rho_{ee} + (i\Delta - \frac{\Gamma}{2})  \tilde \rho_{eg}\label{equ-dm-bloch2-eg}.
\end{align}
\end{subequations}
From~\cref{equ-dm-bloch2-gg,equ-dm-bloch2-ee,equ-dm-bloch2-eg} one can produce the Optical-Bloch equation
\begin{equation}\label{equ-dm-bloch-matrix}
\begin{bmatrix}
    d\rho_{gg}/dt \\
    d\rho_{ee}/dt\\
    d\tilde \rho_{ge}/dt \\
    d\tilde \rho_{eg}/dt
\end{bmatrix}
    =
\begin{bmatrix}
    0 & \Gamma & -\frac{i\Omega^{*}}{2} & \frac{i\Omega}{2} \\
    0 & - \Gamma & \frac{i\Omega^{*}}{2} & - \frac{i\Omega}{2} \\
    - \frac{i\Omega^{*}}{2} & \frac{i\Omega^{*}}{2} & 0 & - i\Delta - \frac{\Gamma}{2}\\    
    \frac{i\Omega^{*}}{2} & - \frac{i\Omega^{*}}{2} & 0 & i\Delta - \frac{\Gamma}{2}
\end{bmatrix}
\begin{bmatrix}
    \rho_{gg} \\
    \rho_{ee}\\
    \tilde \rho_{ge}\\
    \tilde \rho_{eg}
\end{bmatrix}
.
\end{equation}
The electric field amplitude $E_{0}(t)$ should be a profile consistent with the laser pulse of the experiment.
Oreshkina et al.~\cite{Oreshkina2014, Oreshkina2015} use a Gaussian envelope with a constant phase, 
and a Gaussian envelope with a random phase (evaluated with the partial coherent method (PCM)~\cite{Pfeifer2010, Cavaletto2012}). These two cases are considered here, in addition to the case of the homogeneous envelope.

To solve Eqn.~\eqref{equ-dm-bloch-matrix}, it is assumed that initially one hundred percent of the population is in the ground state (i.e., one starts with $\begin{bmatrix}1& 0& 0& 0 \end{bmatrix}$$^{T}$ for the density vector). 
The energy detected from the line emission can be expressed as a function of the detuning parameter
\begin{equation}\label{equ-energy-detect}
    E(\Delta) \propto \Gamma \omega_{0} \int_{-\infty}^{+\infty} \rho_{ee}(t) dt , 
\end{equation}
with $\rho_{ee}(t)$ being evaluated from Eqn.~\eqref{equ-dm-bloch-matrix}. The line intensity is then evaluated from an integral over the detuning parameter:
\begin{equation}\label{equ-line-intensity}
    L = \int E(\Delta) d\Delta .
\end{equation}

Note that the laser pulse parameters are included in the D-M approach via the the electric field ($\vec{E}$), with the pulse envelope imposed on $E_{0}(t)$ and the time dependence of the phase of the electric field included in $\psi(t)$.  The C-R approach includes the intensity profile of the laser via the radiation field density ($\rho(t)$) but does not include the phase of the electric field. The Einstein $A$ and $B$ coefficients are related via the detailed balance relationships and thus the C-R method can be thought of as the limiting case for a perfectly incoherent field.  

As part of this work, codes were developed for both the C-R and D-M methods.  The C-R results have been presented in the literature~\cite{Loch2015}.  Here we show the D-M results for the same conditions as those of Oreshkina et al.~\cite{Oreshkina2014,Oreshkina2015}, to test their conclusions.  Also, in the following section C-R results will be shown which use identical Einstein $A$-coefficients as the D-M calculations and the radiation field densities will also be converted to the equivalent laser intensities.  Note that the two methods should not be expected to produce equivalent results, even for low radiation field densities, as they treat the coherence effects differently.  It is nevertheless interesting to show the results from both approaches, and these are presented in the next section.

\section{Results}
\label{sec:results}

\subsection{LCLS parameter estimation}
The LCLS XFEL parameters for the experiment are described by Bernitt et al.~\cite{Bernitt2012} and previous publications \cite{2010NaPho...4..641E}. The modeling results require the radiation field density parameters (for the C-R results) and the laser intensity parameters (for the D-M results).  
From Bernitt et al.~\cite{Bernitt2012}, the laser pulses vary in duration from 200 to 2000~fs, but mostly within the range of 200--500~fs (G.V.~Brown, private communication). The total energy per laser pulse in the experiment has an upper limit of 3~mJ. 
However the filtering and optical losses after the soft X-ray (SXR) monochromator are expected to reduce the
 total energy per shot to 0.0013--0.39~mJ~\cite{Loch2015}. The LCLS XFEL focal diameter has a range of 3--10~$\mu$m \cite{Heimann2011}.
A value of 10~$\mu$m was chosen for the modeling to make the beam weakly focused. Note that the possibility that the beam had a much larger diameter will be considered later in this paper. These parameters result in a radiation field density ($\rho$) of 
4.62~$\times$~10$^{-7}$ -- 3.46~$\times$~10$^{-4}$~J/m$^{3}$/Hz, and using a laser bandwidth of 1.0~eV the corresponding laser intensity would be in the range 
4.18~$\times$~10$^{12}$ -- 3.14~$\times$~10$^{15}$~W/cm$^{2}$.  Oreshkina et al.~\cite{Oreshkina2014,Oreshkina2015} estimated the laser
intensity to be in the range 10$^{11}$ -- 10$^{14}$~W/cm$^2$.  They used a larger focal diameter than the one given above and also a larger energy per pulse (3~mJ).

The other important characteristic about the LCLS XFEL pulses is their stochastic nature.  Each pulse consists of many short spikes a few fs in duration, with gaps between the spikes also being a few fs long.  The phase during each of the spikes is in general not 
coherent with the previous spikes.  Thus, both the intensity and the phase are stochastic in nature for each pulse.
In the case-studies presented below we first consider the line intensity ratio for individual homogeneous pulses to illustrate the mechanism for the lowering of the line intensity ratio. We then introduce stochastic pulses and evaluate the line ratio for a large number of stochastic pulses to simulate the experimental conditions as closely as possible.

\subsection{C-R model}
The C-R results for these LCLS laser parameters using a number of pulse profiles for $\rho(t)$ are considered first.  Einstein $A$-coefficients of 2.22~$\times$~10$^{13}$~s$^{-1}$ and 6.02~$\times$~10$^{12}$~s$^{-1}$  for the 3C and 3D $A$-values were used, taken from the largest calculation shown in~\cite{Oreshkina2014,Oreshkina2015}. The purpose here is to demonstrate the mechanism for the reduction in the 3C/3D line intensity ratio, with the conclusions being independent of the precise values chosen for the $A$-values.

\subsubsection{Smooth homogeneous pulse}
\begin{figure}
\includegraphics[width=14pc, angle=-90]{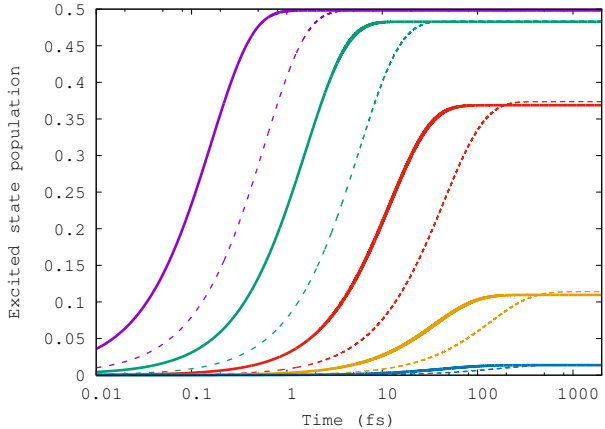}
\caption{\label{fig_cr_pop} Excited state fractional population ($N_e/(N_e+N_g)$) as a function of time for a homogenous 
radiation field density using the C-R method. The solid lines shows the upper level populations for the 3C transition and the dashed lines show
the upper level populations for the 3D transition. Results are shown for laser intensities of 10$^{15}$~W/cm$^{2}$ (purple), 10$^{14}$~W/cm$^{2}$
(green), 10$^{13}$~W/cm$^{2}$ (red), 10$^{12}$~W/cm$^{2}$ (yellow), and 10$^{11}$~W/cm$^{2}$ (blue).}
\end{figure}

Considering first a pulse with a radiation field density that is homogeneous in time, the time-dependent populations can be solved using~\cref{equ-cr-matrix} and the 3C/3D line intensity ratio determined using~\cref{equ-cr-pho}.    
Fig.~\ref{fig_cr_pop} shows the excited states population for the upper levels of the 
3C and 3D transitions for a range of pulse intensities.  Both excited state populations increase towards a constant (steady-state) value during the homogeneous pulse. However, due to the different 
Einstein $A$ coefficients for the 3C and 3D transitions, the two excited states converge onto this value at different rates. The excited state population for the upper level of the 3C line reaches steady-state in a shorter time than the corresponding 3D population. For low radiation field densities the steady-state population value depends linearly on the radiation field density and results in an excited state population fraction that is less than 0.5. As the radiation 
field density increases, the excited states reach their steady-state value in a much shorter time and the steady-state value is no longer directly proportional to the radiation field density.
It can also be seen that the maximum value for the steady-state excited population fraction is 
0.5, the high radiation field density limit for the excited population in the C-R method. In this case, the populating and 
depopulating of the excited states happen simultaneously, in other words the process is always incoherent,
which leads to steady and non-oscillating excited state populations. 

\begin{figure}
\includegraphics[width=14pc, angle=-90]{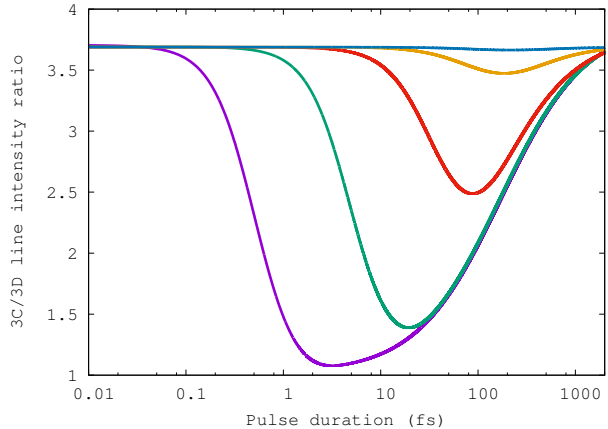}
\caption{\label{fig_cr_ratio}  The 3C/3D line intensity ratio as a function of pulse duration for a homogenous 
radiation field density using the C-R method. Results are shown for laser intensities of
10$^{15}$~W/cm$^{2}$ (solid purple line), 10$^{14}$~W/cm$^{2}$ (solid green line),
10$^{13}$~W/cm$^{2}$ (solid red line), 10$^{12}$~W/cm$^{2}$ (solid yellow line),
10$^{11}$~W/cm$^{2}$ (solid blue line).}
\label{fig:crhom3c3d}
\end{figure}

The 3C/3D line intensity ratio for a homogenous radiation field density is shown in Fig.~\ref{fig:crhom3c3d}. For laser intensities above approximately 1~$\times$~10$^{12}$~W/cm$^2$ there is a reduction in the line intensity ratio below the oscillator strength ratio value. The reduction was shown previously~\cite{Loch2015} to be primarily due to contributions to the emission during the XFEL interaction with the plasma being different from the contribution after the pulse has left the plasma volume.  For the intense pulses, the 3D intensity always has a larger fraction of its emission coming from this 'after the pulse' component than the 3C intensity.  This results in a reduction in the line intensity ratio below the oscillator strength ratio value.

\subsubsection{Stochastic pulse}
Consider next the C-R results for a stochastic profile of the pulse. We generate a random set of Gaussian profiles, each with 0.2~fs standard deviation and remove a random number of Gaussians to produce a pulse profile similar to that shown on the LCLS web page, see Fig. 4 of Loch et al.~\cite{Loch2015}.  We normalize the stochastic pulse profile so that the integrated intensity is equivalent to a homogeneous radiation field density.  We then use this value to label the stochastic pulse, which allows us to compare the two sets of results. 

\begin{figure}
\includegraphics[width=14pc, angle=-90]{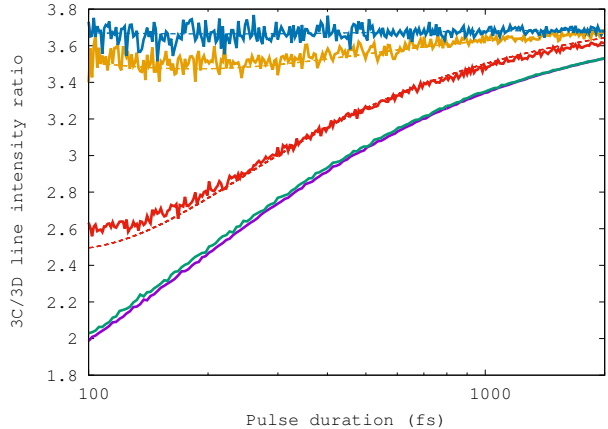}
\caption{\label{fig_cr_ratio_stoch} {\color{red} C-R values} for the 3C/3D line intensity ratio as a function of pulse duration.  The stochastic results take an average of 80 stochastic 
pulses for each data point. The homogeneous results are the same as those shown in Fig.~\ref{fig:crhom3c3d}. The solid lines show the stochastic results and the dashed lines show the homogeneous data.  Results are shown for intensities of
10$^{15}$~W/cm$^{2}$ (purple), 10$^{14}$~W/cm$^{2}$ (green), 10$^{13}$~W/cm$^{2}$ (red), 10$^{12}$~W/cm$^{2}$ (yellow), 10$^{11}$~W/cm$^{2}$ (blue).}
\end{figure}

Fig.~\ref{fig_cr_ratio_stoch} shows the comparison of line ratio using the C-R method with both the homogeneous and stochastic pulses. 
The stochastic features of the pulse profiles do not change the overall trend of the line ratio using the C-R model.  This is because the stochastic laser intensity spikes have only small (i.e., a few fs) gaps between them.  Thus, for intense pulses the excited populations are still driven close to their steady-state values and do not have time to decay significantly during the gap between the spikes.  In the stochastic simulations we use different pulses for the 3C and 3D transitions, and have many pulses for each set of pulse parameters.  Each point in Fig.~\ref{fig_cr_ratio_stoch} was generated using 80 stochastic pulse profiles for the 3C and 80 pulses for the 3D.  Note the stochastic pulse simulations produce a similar reduction in the line ratio to that obtained from the homogeneous pulse calculations, i.e.  
the 3C/3D line ratios are lower for shorter and intense pulses.  Note that the experiment would have involved a large number of pulses of different intensities and pulse durations. If the distribution of pulse conditions was known, then it would be possible to compare with a simulated line ratio for the same set of pulse distributions. Such a simulation could be used to explore the sensitivity to the $A$-values employed in the model, resulting in a recommended range of values on the $A$-value ratio. While the experimental distribution of pulse conditions is not currently known well enough to perform such a comparison, it should be pointed out that the C-R model implies that pulse intensities above 10$^{12}$~W/cm$^2$ are required to produce a reduction in the line ratio.

\subsection{D-M model}
We next consider the D-M approach for different pulse envelopes. The same laser bandwidth (1.0~eV) and $A$-values are used as those chosen by Oreshkina et al.~\cite{Oreshkina2014,Oreshkina2015}, to allow a direct comparison to be made with their results.  As in the discussion of the C-R results, the conclusions that are drawn here will be general and not dependent upon the specific values chosen for the $A$-values for Fe$^{16+}$.  

\subsubsection{Smooth homogeneous pulse}
In the D-M approach, the level populating and depopulating mechanisms are slightly different 
from the C-R model, as the process involves an intermediate step which contains two polarization states, 
$\rho_{ge}$ and $\rho_{eg}$. This characteristic enables the Rabi-oscillation of the populations and is required for intense radiation fields and coherent systems. 

\begin{figure}[h]
\begin{minipage}{14pc}\label{fig_dm_pop_flat_smooth_1}
\includegraphics[width=10pc, angle=-90]{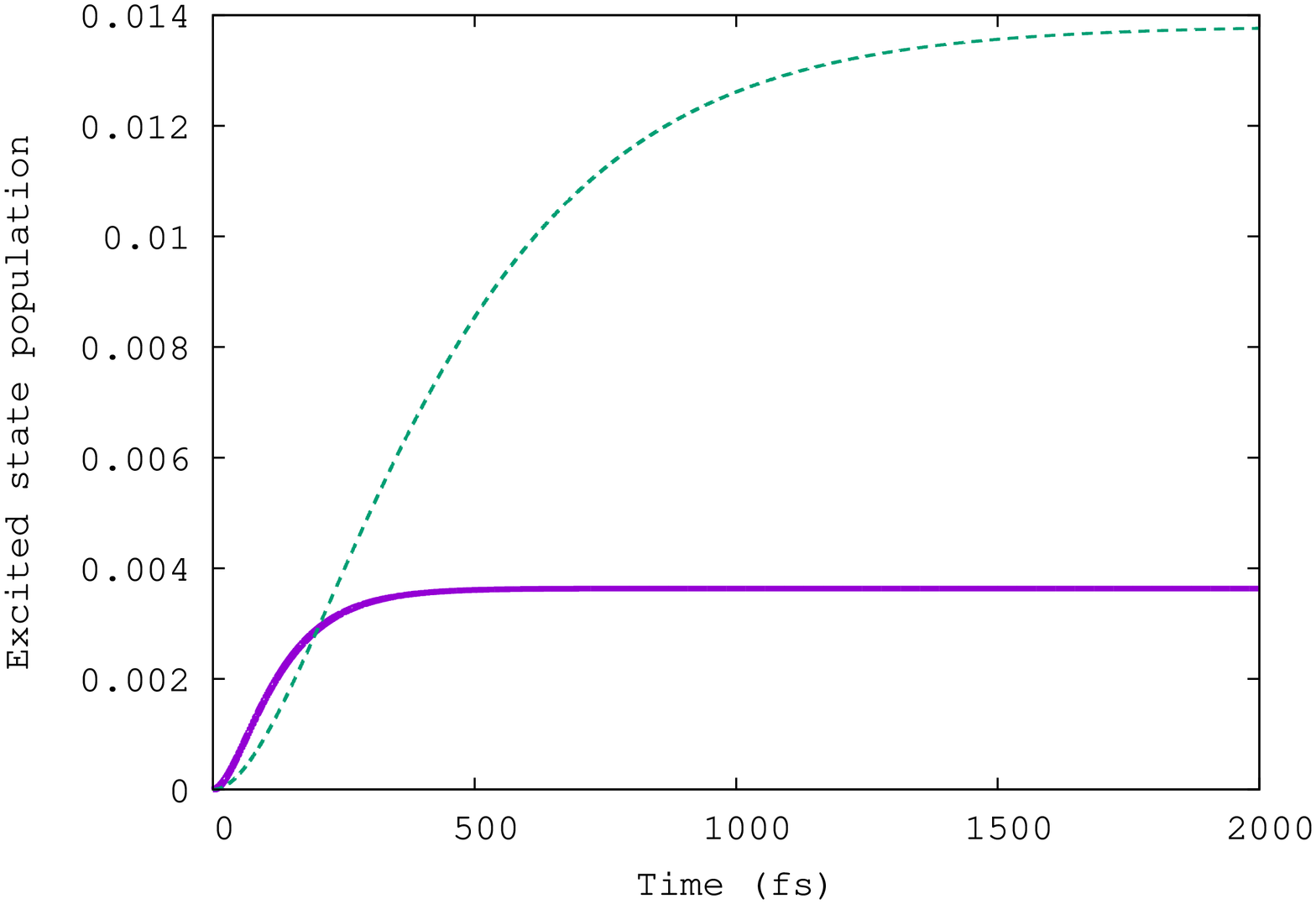}
\end{minipage}\hspace{2pc}
\begin{minipage}{14pc}\label{fig_dm_pop_flat_smooth_2}
\includegraphics[width=10pc, angle=-90]{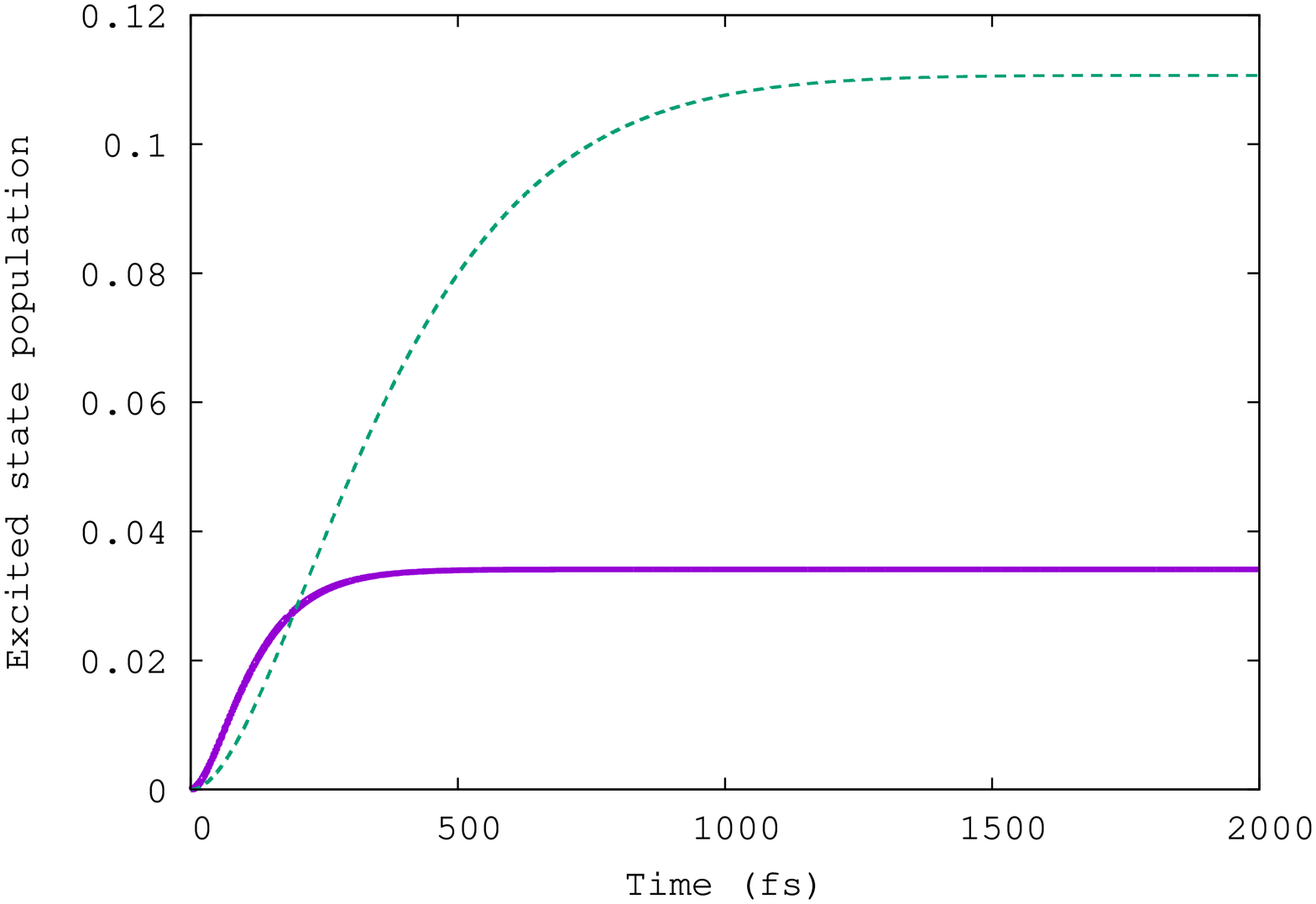}
\end{minipage}\\

\begin{minipage}{14pc}\label{fig_dm_pop_flat_smooth_3}
\includegraphics[width=10pc, angle=-90]{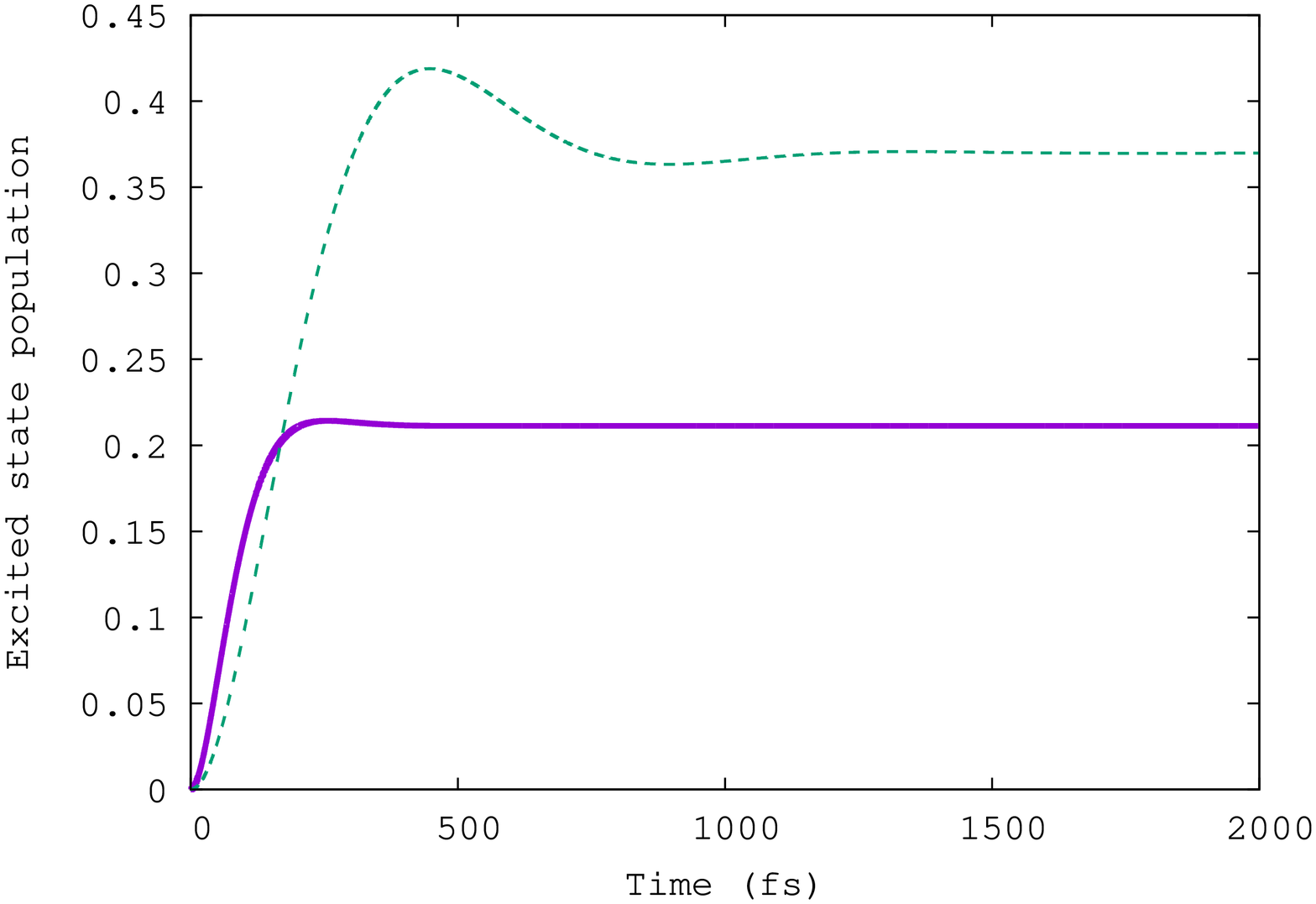}
\end{minipage}\hspace{2pc}
\begin{minipage}{14pc}\label{fig_dm_pop_flat_smooth_4}
\includegraphics[width=10pc, angle=-90]{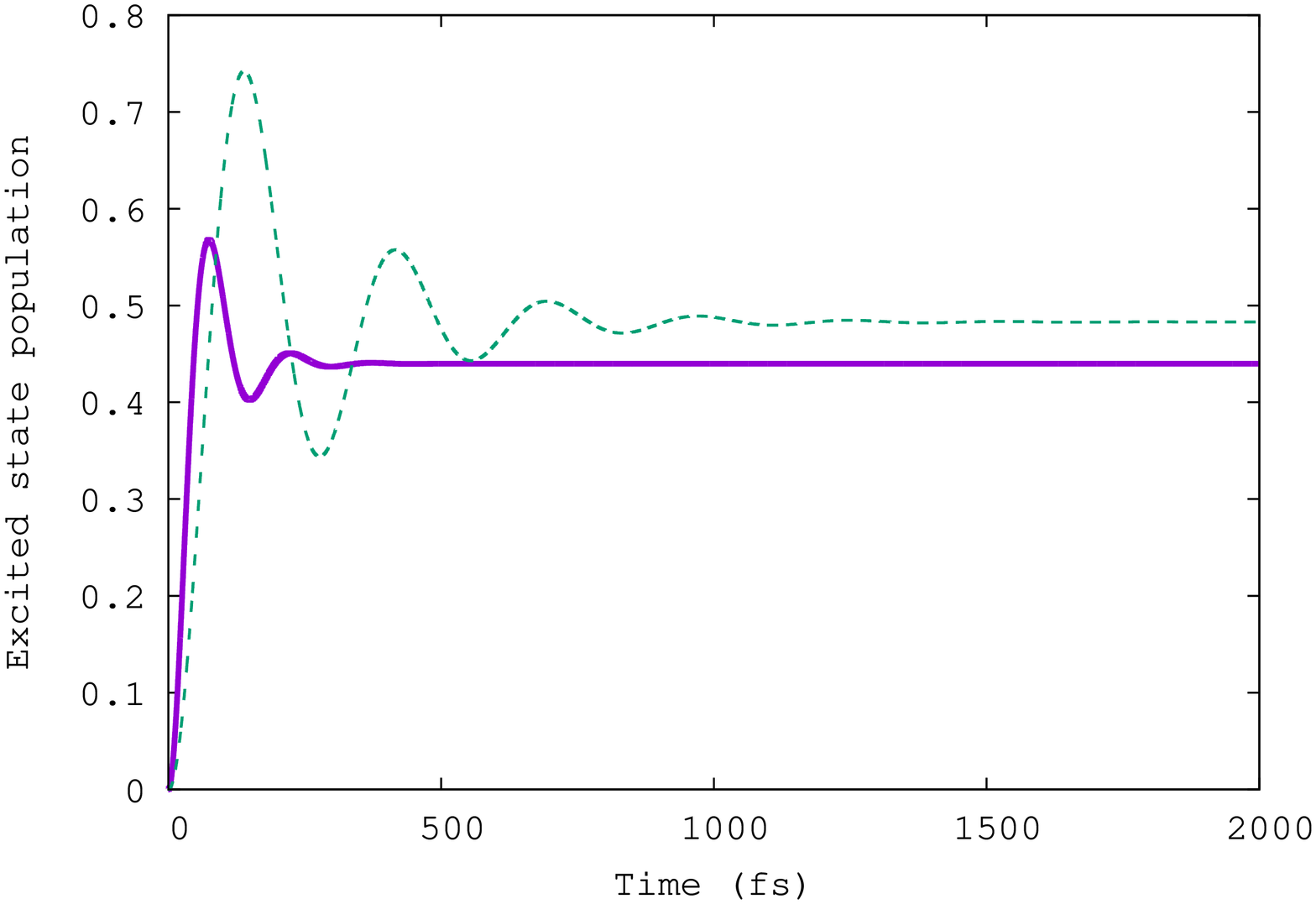}
\end{minipage}\\

\begin{minipage}{14pc}\label{fig_dm_pop_flat_smooth_5}
\includegraphics[width=10pc, angle=-90]{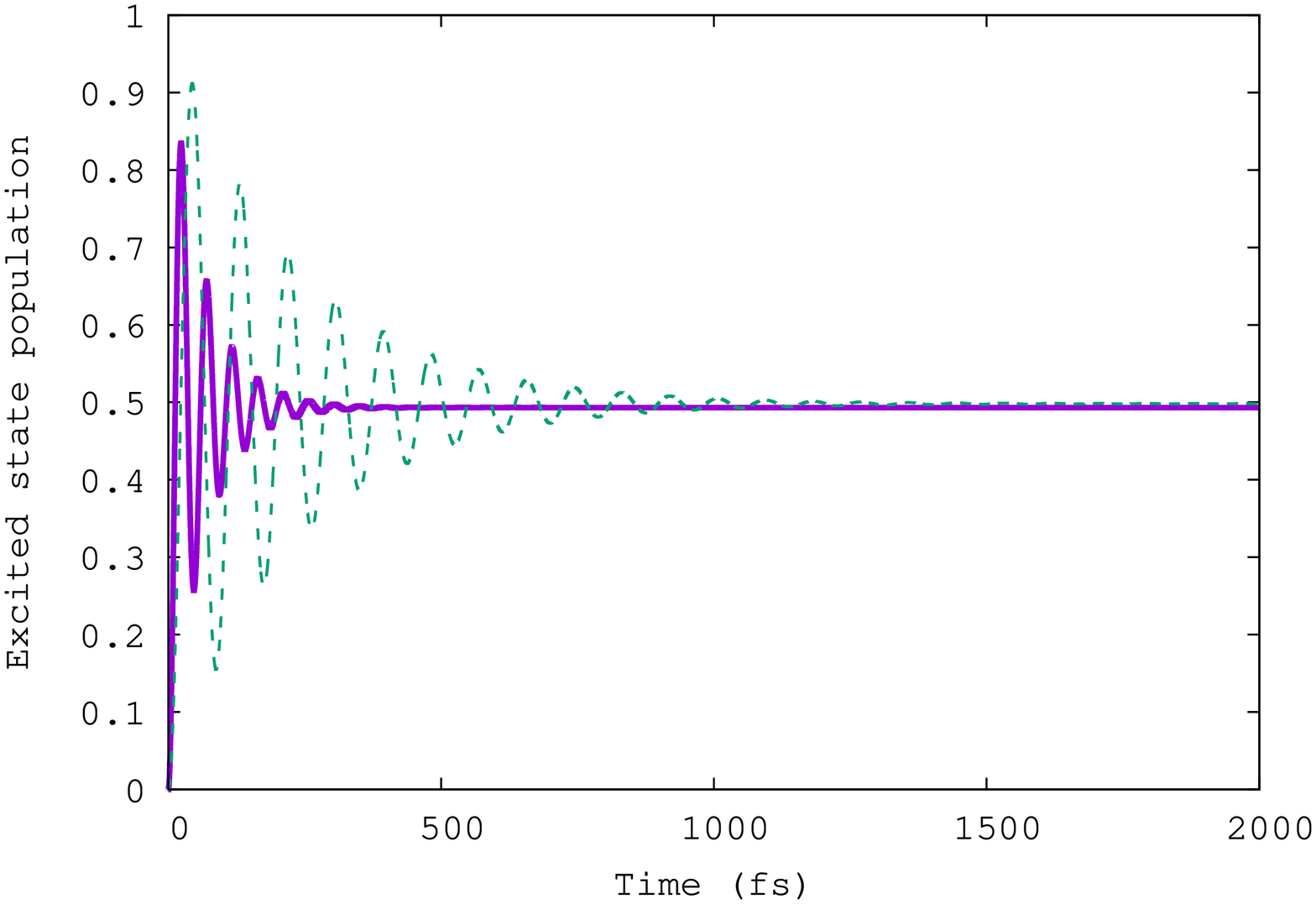}
\end{minipage}\hspace{2pc}
\begin{minipage}{14pc}\label{fig_dm_pop_flat_smooth_6}
\includegraphics[width=10pc, angle=-90]{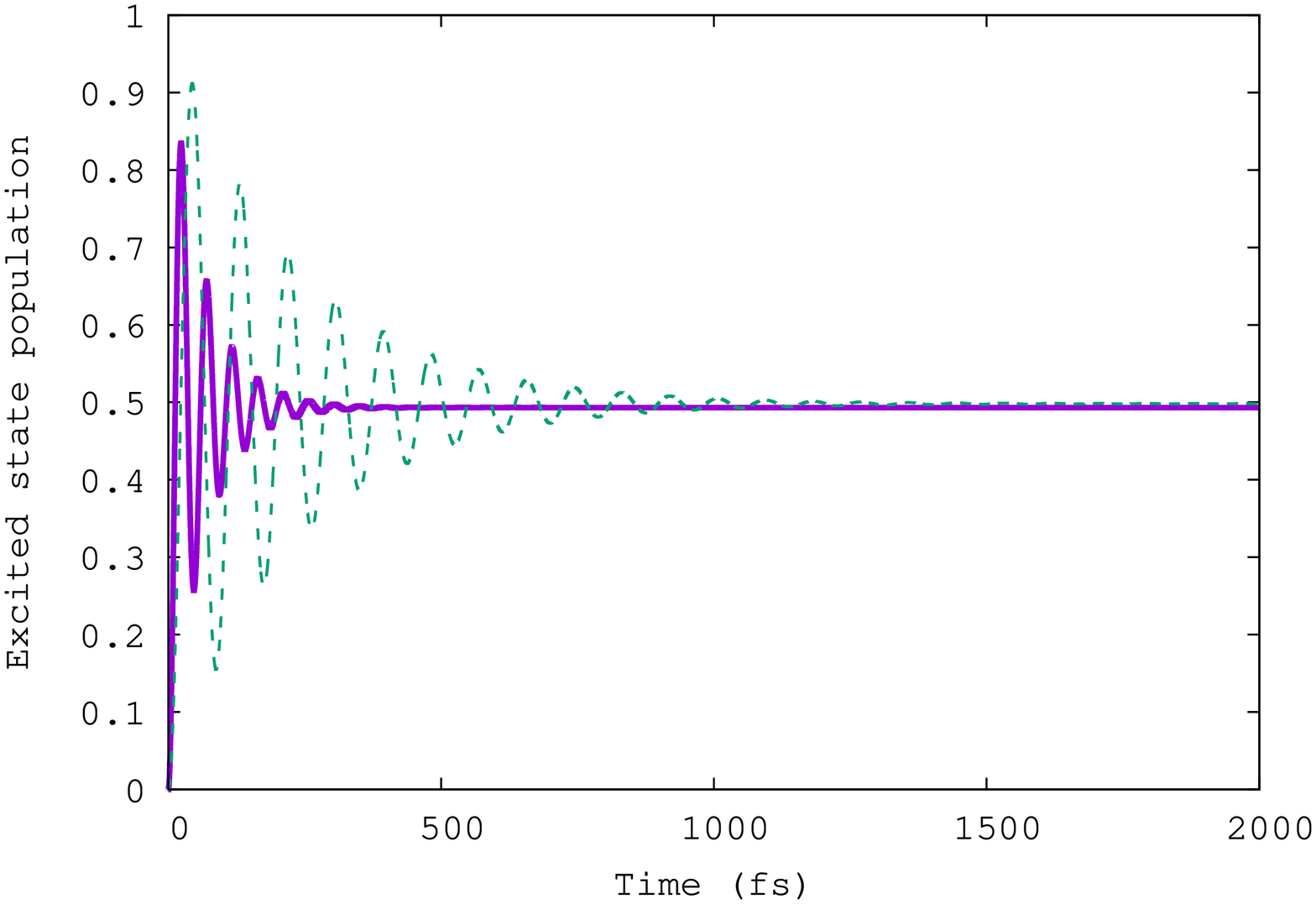}
\end{minipage}\\

\caption{\label{fig_dm_pop_flat_smooth}  Excited state fractional populations as a function of time under a continuous flat 
pulse using the D-M approach. The solid purple lines show the excited 3C populations and the dashed green lines show the excited 3D populations.  Results are shown for 10$^{9}$~W/cm$^{2}$ (row 1, column 1), 10$^{10}$~W/cm$^{2}$ (row 1, column 2),  10$^{11}$~W/cm$^{2}$ (row 2, column 1), 10$^{12}$~W/cm$^{2}$ (row 2, column 2), 10$^{13}$~W/cm$^{2}$ (row 3, column 1), 10$^{14}$~W/cm$^{2}$ (row 3, column 2).}
\end{figure}

We consider first a homogeneous pulse, that is $E_{0}(t)$ is a constant in time, with the value determined from the laser intensity.~\cref{equ-dm-bloch-matrix} is used to evaluate the time-dependent populations  and~\cref{equ-line-intensity} is used to evaluate the Fe XVII 3C/3D line intensity ratio.
Fig.~\ref{fig_dm_pop_flat_smooth} shows the excited state populations as a function of time using the D-M approach for a range of homogeneous pulse intensities. For low intensities the populations increase smoothly to a steady-state value, with a similar shape to the C-R results.  There is, however, a noticeable difference: the steady-state value can be different for the two transitions.  It is still the case that the 3C excited population reaches steady-state in a shorter time than the 3D excited population.
At higher intensities ($\sim10^{11}$~W/cm$^{2}$ and above), Rabi-flopping starts to become apparent in both the 3C and 3D populations.  Thus, the duration of the pulse can make a large difference in the relative emission for the two lines.  One pulse could result in a 3C excited population that is greater than the 3D excited population, while a slightly longer pulse could lead to the opposite.  It can also be seen that for the D-M method for coherent pulses, the 3C/3D line ratio could be  higher than or smaller than the oscillator strength ratio, depending upon the relative populations of the two excited states.
This will be shown in more detail in the next section.

\subsubsection{Coherent Gaussian pulse}
\begin{figure}[h]
\begin{minipage}{14pc}
\includegraphics[width=10pc, angle=-90]{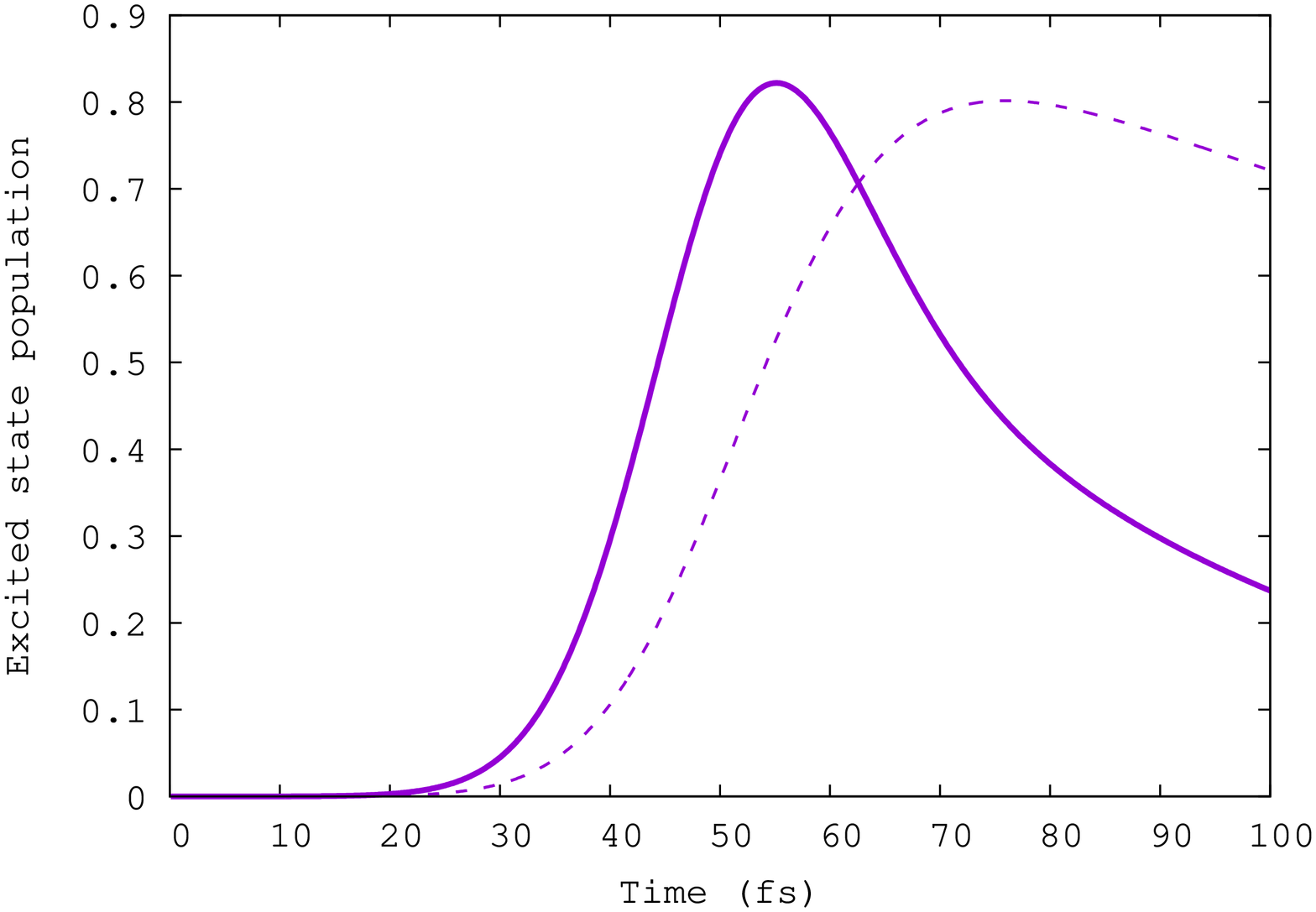}
\end{minipage}\hspace{2pc}
\begin{minipage}{14pc}
\includegraphics[width=10pc, angle=-90]{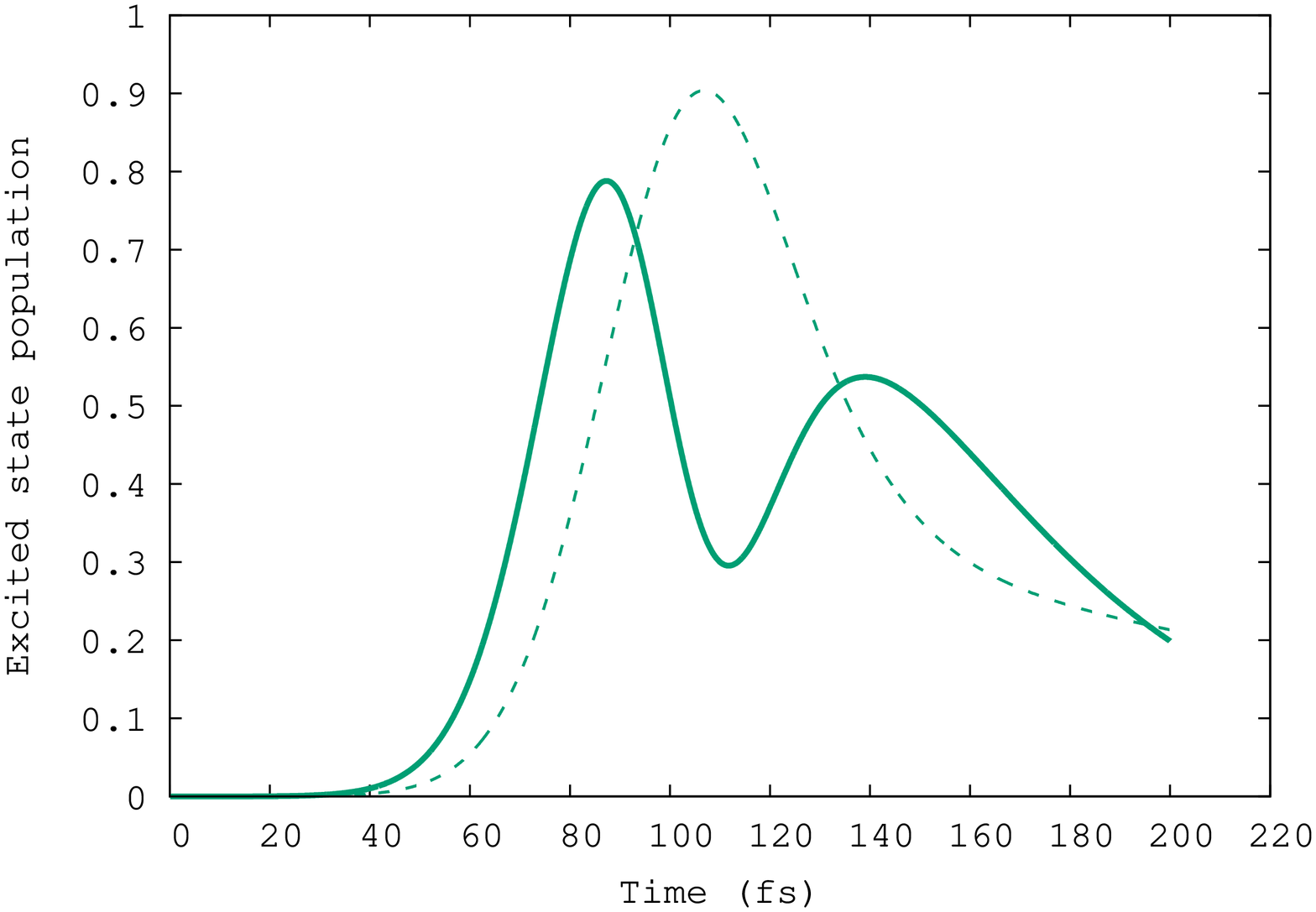}
\end{minipage}
\caption{\label{fig_dm_pop_gau}  Excited state fractional populations as a function of time for a Gaussian 
pulse with intensity 10$^{13}$~W/cm$^{2}$ using the D-M model.
The left panel displays the 100-fs results: the solid (purple) line indicates the 3C population and the dashed (purple) line indicates the 3D population. The right panel displays the 200-fs results: the solid (green) line indicates the 3C population and the dashed (green) line indicates the 3D population.}
\end{figure}

Oreshkina et al.~\cite{Oreshkina2014, Oreshkina2015} modeled the Fe XVII experiment using a D-M approach with a Gaussian profile as the pulse envelope. 
We consider the same case here, to allow us to compare our D-M results with theirs.
We start with pulses which have a coherent phase for the duration of the pulse ($\psi(t)=0$).
Fig.~\ref{fig_dm_pop_gau} shows the time evolution of the excited population fractions for a pulse with intensity of 1~$\times$~10$^{13}$~W/cm$^2$ and two different pulse durations (100~fs and 200~fs), showing characteristic Rabi-flopping. The Rabi-frequency of the 3C populations is more rapid than the 3D, due to the larger $A$-value for the 3C transition.
This difference in Rabi-frequency can result in quite different excited populations at the end of the laser pulse interaction with the plasma. Considering these two pulse durations as an illustrative example:
for the 100~fs case, the 3D transition has a much larger excited population at the end of the pulse than the 3C excited population, while for the 200~fs case the two have almost the same population fraction. This behavior drives the 3C/3D line intensity
ratio for the 100~fs case to be much smaller than the oscillator strength value.  For these coherent and intense laser conditions, the line intensity ratio produced from these populations would not necessarily be equivalent to the oscillator strength ratio.  Furthermore, the contribution to the emission from the time after the laser pulse has left the plasma volume is quite sensitive to the population in the excited state at the end of the laser pulse. Again one has the scenario where the
emission from the `after-the-pulse' component will be quite different in the two cases, producing quite different line ratio values for these two pulses. 

\begin{figure}
\includegraphics[width=14pc, angle=-90]{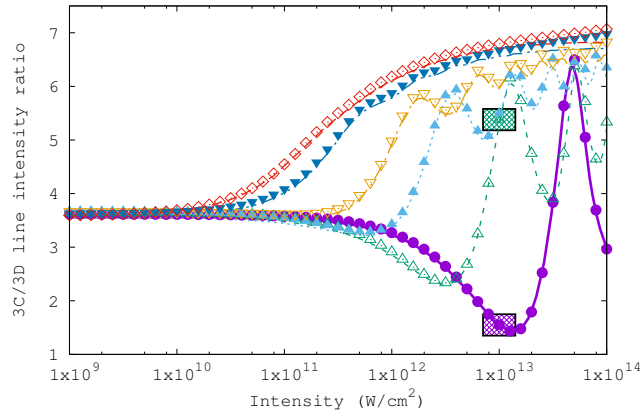}
\caption{\label{fig_dm_ratio_gau} 3C/3D line intensity ratio as a function of radiation field intensity under a Gaussian 
pulse using the D-M model compared with Oreshkina el al.~\cite{Oreshkina2014, Oreshkina2015}.  In all cases the symbols show the results from the work of this paper and the lines show the results of Oreshkina et al.~\cite{Oreshkina2015}.
Results are shown for 100~fs (purple), 200~fs (green), 400~fs (blue),  600~fs (yellow), 1200~fs (dark blue), and 2000~fs (red).}
\end{figure}

Fig.~\ref{fig_dm_ratio_gau} shows the 3C/3D line ratio as a function of pulse duration 
for coherent Gaussian pulses. We obtain very similar line ratio results to those of Oreshkina et al.~\cite{Oreshkina2014, Oreshkina2015}. It is useful to consider the two pulse durations shown in Fig.~\ref{fig_dm_pop_gau}.
The 3C/3D line ratios for the two scenarios shown in Fig.~\ref{fig_dm_pop_gau} are shown by the purple and green squares in Fig.~\ref{fig_dm_ratio_gau}. 
For the 100-fs pulse (where the 3D population fraction is greater than the 3C value at the end of the pulse), the line ratio is 1.55 which is much smaller than the 3C/3D oscillator strength ratio, as one might expect from the populations.
For the 200~fs pulse (where the 3D population fraction is about the same as the 3C at the end of the pulse), the ratio is 5.38. 
Fig.~\ref{fig_dm_ratio_gau} also shows that for coherent pulses a change in the line ratio from the oscillator strength ratio requires pulse intensities above about 1~$\times$~10$^{11}$~W/cm$^2$.

\subsubsection{Stochastic Gaussian pulse}
\begin{figure}
\includegraphics[width=14pc, angle=-90]{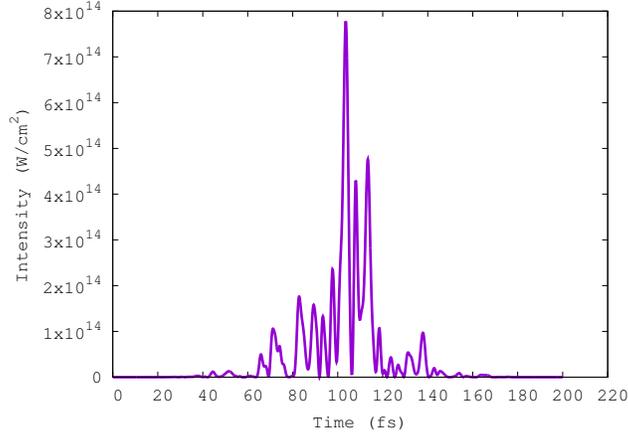}
\caption{\label{fig_dm_pulse_stoch_gauss} A sample stochastic pulse with Gaussian envelope for a 200 fs pulse duration.}
\end{figure}

\begin{figure}
\includegraphics[width=14pc, angle=-90]{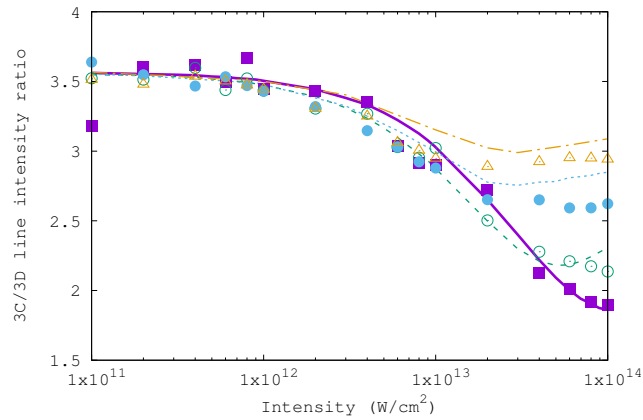}
\caption{\label{fig_dm_ratio_stoch} The 3C/3D line intensity ratio as a function of radiation field intensity for a stochastic 
Gaussian pulse using the D-M model. The symbols show the current results and the lines show the results of
Oreshkina et al.~\cite{Oreshkina2014,Oreshkina2015}. Results are shown for 100~fs (purple), 200~fs (green), 400~fs (blue), and 600~fs (yellow).}
\end{figure}

To model the LCLS pulse parameters more accurately, the stochastic features of the pulse need to be included. We use the PCM~\cite{Pfeifer2010, Cavaletto2012} to model the stochastic nature of the pulse intensity and phase. Fig.~\ref{fig_dm_pulse_stoch_gauss} shows a stochastic pulse intensity
generated using the PCM. Note that it still has a Gaussian envelope, but there are now many stochastic spikes of intensity throughout the pulse. Note also that the electric field strength and the phase are both stochastic and complex. These stochastic pulses can now be modeled using the D-M formalism to produce a 3C/3D line intensity ratio.
Fig.~\ref{fig_dm_ratio_stoch} shows the comparison of the calculated 3C/3D line intensity ratio with the results of Oreshkina et al.~\cite{Oreshkina2014, Oreshkina2015}. 
The line ratio results are calculated from an average of 80 pulses using a bandwidth of 1~eV, and the results are in good agreement with 
Oreshkina et al.~\cite{Oreshkina2014,Oreshkina2015}. We were, however, not able to achieve convergence within~10~or~20~pulses as stated in their paper; in general it took more runs to achieve convergence on the average line ratio value.
The calculated line ratios are all below the oscillator strength ratio for intensities above~$\sim$~10$^{12}$~W/cm$^{2}$. The bandwidth 
of the pulse also affects the coherence of the pulse and the duration of the spikes in the intensity, thus it strongly affects the line ratio. If the bandwidth is very small, then the pulse profile becomes much more coherent and the spikes in intensity are wide.  In this limit the stochastic pulses produce line ratio values very close to the
coherent Gaussian pulses from Fig.~\ref{fig_dm_ratio_gau}.

\begin{figure}
\includegraphics[width=30pc, angle=0]{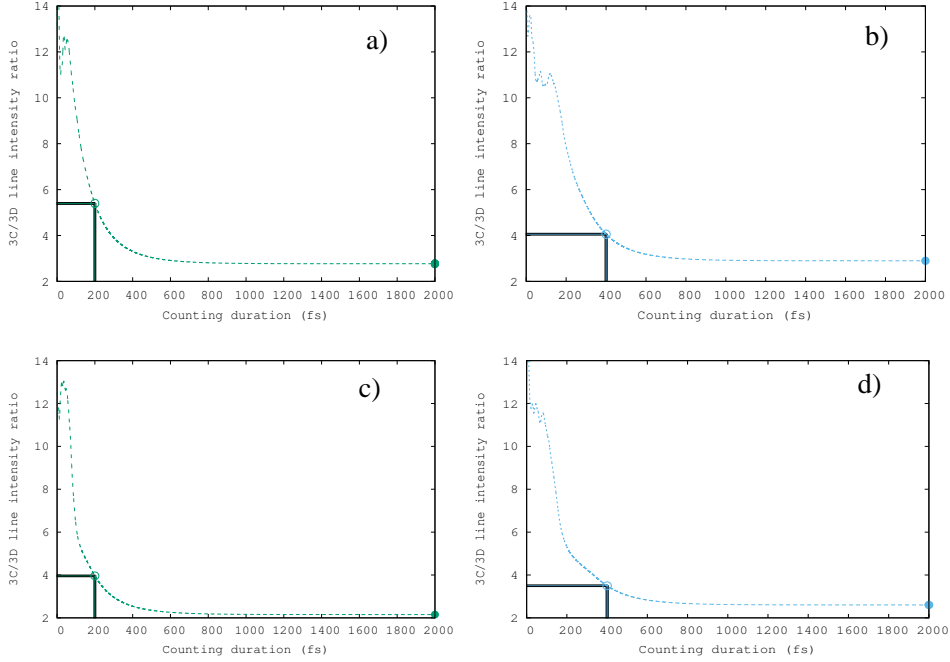}
\caption{\label{fig_dm_after_pulse} The measured line intensity ratio as a function of counting duration. D-M results are shown for Gaussian pulse envelopes with the following conditions: a) with 200~fs and an intensity of 1~$\times$~10$^{13}$~W/cm$^{2}$, b) 400~fs and an intensity of 1~$\times$~10$^{13}$~W/cm$^{2}$, c) 200~fs and an intensity of 1~$\times$~10$^{14}$~W/cm$^{2}$, d) and 400~fs and an intensity of 1~$\times$~10$^{14}$~W/cm$^{2}$. The hollow circle shows what the value would be if one stopped counting photons after the time specified by the Gaussian width and the solid circles show the results if one kept integrating until the final time.}
\end{figure}

It should also be noted that the emission from the plasma after the pulse has left the plasma volume is still a strong factor in lowering the line intensity ratio below the oscillator strength value.
In the D-M approach using Gaussian envelopes for the pulses, it is difficult to define a before- and after-the-pulse component to the emission as the Gaussian envelope will continue far beyond the defined width of the pulse. However, with the laser intensity dropping off, one would expect the emission characteristics at later times to be quite different from the emission when the pulse is at its peak intensity. Figure ~\ref{fig_dm_after_pulse} shows what the measured line intensity ratio would be if one stopped counting photons at different times, for 4 different pulse profiles. This was generated using the D-M code, with 80 stochastic pulses per datapoint, a Gaussian envelope of either 200 or 400 fs  width, and intensities of 1~$\times$~10$^{13}$~W/cm$^{2}$ and 1~$\times$~10$^{14}$~W/cm$^{2}$. It can be seen that the contribution from the emission after the pulse has finished its strongest interaction with the plasma is an important factor in producing a 3C/3D line intensity ratio that is lower than the oscillator strength value. In fact, without this contribution in the D-M approach the results would often be above the oscillator strength ratio. Thus, for both the C-R and D-M approaches it is important to keep counting the emission beyond the main interaction phase of the laser with the plasma.



As a final illustration of the results using the D-M approach, a simulation was carried out for a distribution of pulse intensities and pulse durations. Using a laser bandwidth of 1.0~eV, a distribution of pulse intensities, with 10 evenly spaced points per decade from 10$^{11}$ to 10$^{14}$~W/cm$^{2}$, and a distribution of linearly spaced pulse durations ranging from~200~to~500~fs, a total line intensity for the 3C and 3D lines was produced. The two total line intensities were then used to produce a 3C/3D line intensity ratio, giving a value of  2.71. It should be noted that the pulse parameters and distributions are not well known from the experiment, so this type of investigation should not be considered to be a true simulation of the experiment, but an illustration that pulse parameters in this range of intensities and durations can produce a line intensity ratio close to the value that was measured.  For the $A$-values chosen for this simulation, some pulse intensities at (or above) 10$^{13}$~W/cm$^2$ are required to produce line ratios in the range measured by the experiment. It would clearly be very useful to be able to use the observed line intensity ratio, and knowledge of the pulse parameters, to determine what the 3C/3D A-value ratio would need to be to produce agreement with the experiment (i.e., to make no assumption about the A-values for either line, but to determine the ratio from the experiment).  However, without more accurate knowledge of the pulse parameters, this does not currently appear to be possible.  The next section explores this concept in more detail.

\subsubsection{Photon counts}
If the laser intensity is significantly below 10$^{11}$ W/cm$^{2}$, one would expect the line intensity ratio 
to be close to the oscillator strength ratio. In recent discussion with the experimentalists, it was pointed out to us that the 
defocusing of the laser would produce a beam much more weakly focused than we assumed in our model.  While we had assumed a beam radius of 5~$\mu$m, it was likely to be closer to 0.5 mm (FWHM), i.e. a factor of 100~times wider.  This change would produce intensities a factor of 10$^4$ weaker, so the range of pulse intensities would be 4.18~$\times$~10$^{8}$ -- 3.14~$\times$~10$^{11}$~W/cm$^{2}$.  In this range, the measured line intensity ratio would be expected to be the same as the
oscillator strength ratio. 

\begin{figure}
\includegraphics[width=14pc, angle=-90]{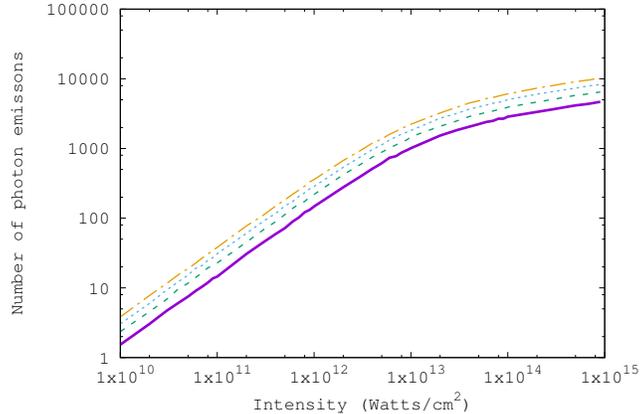}
\caption{\label{fig_dm_photon_stoch} Averaged photon counts for the 3C line as a function of radiation field intensity
 for stochastic Gaussian pulses using the D-M model. Results are shown for 200~fs (solid purple line), 300~fs (dashed green line), 400~fs (dotted blue line), and 500~fs (dot-dashed yellow line).}
\end{figure}

It is instructive to consider the photon counts produced from each pulse, remembering that the LCLS experiment consisted of a large number of individual pulses, with the final line intensity being the result from all of the pulses combined. 
Fig.~\ref{fig_dm_photon_stoch} shows the photon emission as a function of pulse intensity. The weak pulses produce only a few photons, and the number of photons produced increases linearly with pulse intensity until about 10$^{12}$~W/cm$^2$. Thus, 
the more intense pulses produce more photons from the plasma.  For the line intensity ratio to be dominated by the pulse intensities in the 4.18~$\times$~10$^{8}$ -- 3.14~$\times$~10$^{11}$~W/cm$^{2}$ range, it would be very important that no pulses had intensities above this range.  It would only take a few pulses above 10$^{13}$~W/cm$^2$ for those pulses to dominate the line intensities, and hence the line ratio.  This topic will be explored in future work.  It would also be of great benefit if an experiment could be performed where no pulses with intensities above $\sim$~10$^{12}$~W/cm$^2$ were allowed to interact with the plasma.  In such an experiment, the observed line intensity ratio is expected to be a good indication of the 3C/3D oscillator strength ratio.

\section{Conclusions}
\label{sec:conclusion}

A review has been presented of two time-dependent methods that have been used to model the Fe XVII 3C/3D line intensity ratio for an intense laser field, 
the C-R and D-M approaches. Both methods show a reduction in the line intensity ratio below the oscillator strength ratio for pulses with intensities above $\sim$~10$^{12}$~W/cm$^2$. A significant factor in lowering the line intensity ratio for both methods is the contribution to the emission from the plasma after the laser pulse has left the plasma volume.
We confirm the importance of the effects previously reported by Oreshkina et al.~\cite{Oreshkina2014,Oreshkina2015}: the non-linear effects in the D-M method and the stochastic nature of the laser pulses.
As stated earlier, it is likely that the majority of the FEL X-ray pulse intensities in the experiments presented by Bernitt~et~al. are below 1~$\times$~10$^{12}$~W/cm$^{2}$. Since the presence of even a small number of pulses above this threshold could lower the observed 3C/3D line intensity ratio below the oscillator strength ratio, an experiment which could ensure there were no pulses above this threshold, and with well constrained pulse parameters, would allow a conclusive statement about the 3C/3D oscillator strength ratio to be made.

\begin{acknowledgments}
Computational work was carried out at the High Performance Computing Center (HLRS) in Stuttgart, Germany, and on a local cluster at Auburn University.
Ye Li would like to thank Dr.~Uwe~Konopka for helpful translation and experimental advice.
\end{acknowledgments}
\bibliography{asos_fe17_short_v3_1}

\end{document}